\documentclass[aps,prl,twocolumn,showpacs,superscriptaddress,floatfix,amssymb,amsmath]{revtex4}

\usepackage{hyperref}
\usepackage{graphicx}
\usepackage{epstopdf}
\usepackage{color}
\usepackage{bm}

\begin{document}

\title{Frequency combs with weakly lasing exciton-polariton condensates}
\author{K. Rayanov}
 \affiliation{New Zealand Institute for Advanced Study, Centre for Theoretical Chemistry
 \& Physics, Massey University, 0745 Auckland, New Zealand}
\author{B.~L.~Altshuler}
\affiliation{New Zealand Institute for Advanced Study, Centre for Theoretical Chemistry
 \& Physics, Massey University,  0745 Auckland, New Zealand}
\affiliation{Physics Department, Columbia University, New York, USA}
\author{Y.~G.~Rubo}
 \affiliation{Instituto de Energ\'{\i}as Renovables, Universidad Nacional Aut\'onoma de M\'exico,
 Temixco, Morelos, 62580, Mexico}
\author{S.~Flach}
 \affiliation{New Zealand Institute for Advanced Study, Centre for Theoretical Chemistry
 \& Physics, Massey University,  0745 Auckland, New Zealand}
\affiliation{Center for Theoretial Physics of Complex Systems, Institute for Basic Science, Daejeon, Korea}

\date{\today}

\begin{abstract}
We predict the spontaneous modulated emission from a pair of exciton-polariton condensates
due to coherent (Josephson) and dissipative
coupling. We show that strong polariton-polariton interaction generates
complex dynamics in the weak-lasing domain way beyond Hopf bifurcations. As a
result, the exciton-polariton condensates exhibit self-induced oscillations and emit an
equidistant frequency comb light spectrum. A plethora of possible emission spectra with
asymmetric peak distributions appears due to spontaneously broken time-reversal
symmetry. The lasing dynamics is affected by the shot noise arising from the influx of
polaritons. That results in a
complex inhomogeneous line broadening.
\end{abstract}


\maketitle

Condensation of exciton-polaritons (EP's) in semiconductor
microcavities formed by two distributed Bragg mirrors with quantum wells between them has
been experimentally observed
\cite{Kasprzak2006,Balili2007,Lai2007,Baumberg2008,Wertz2009}. Being incoherently excited
in the microcavity, EP condensates are in general out of thermodynamic equilibrium. EP condensates
refuel their particle depot through absorption of cavity photons and emit coherent
light due to tunneling of the composite EP states through distributed Bragg mirrors.
Sample inhomogeneity, either accidental or intentional,
can induce several condensation centers
(CC's) \cite{Lai2007,Baas2008,Krizhanovskii2009,Kim2011}. At low enough pumping, one
expects a system of disconnected BEC droplets emitting light at different uncorrelated
frequencies. As the pumping increases the condensates tend to establish mutual coherence and emit
in a laser mode \cite{Baas2008}. Already two CCs can synchronize and emit at a single
joint frequency \cite{Wouters2008,Aleiner2012}. This is possible because the condensates
exchange particles due to Josephson coupling and adjust their emission
frequencies, which in turn depend on the number of condensed particles due to the
polariton-polariton repulsion. In addition to the coherent Josephson coupling, there can be a
dissipative (radiative) coupling between CC's, which reflects the dependence of the
losses in the system on the symmetry of singe-particle states. That new stationary regime called 
weak lasing takes place when pumping rates reside between some minimal and
maximal rates of losses \cite{Aleiner2012}. In the weak lasing regime, the system
is stabilized by the formation of specific many-particle states which adjust the balance
between
gain and loss in the system. 

In this Letter we show that in the weak lasing regime two CCs can emit not only at a
\emph{single} frequency, but also at a whole {\em frequency comb} which in principle
contains an infinite number of equidistant lines of coherent lase-like radiation. This emission reflects the fact of
formation of spontaneous selfsustained anharmonic oscillations of both the occupation numbers and the
relative phase between the condensates, in sharp contrast to previously reported damped Josephson
oscillations \cite{Sarchi2008,Shelykh2008,Lagoudakis2010}.
We study possible emission spectra and the way 
%as function of pumping and the system
%parameters, and demonstrate how 
they are affected by noise. While the emission frequency of
single-line EP lasers resides in the eV range
\cite{Kasprzak2006,Lai2007,Krizhanovskii2009,Baas2008,Love2008}, the modulation frequency
of comb emission can be adjusted to be in the terahertz and sub-terahertz range.
%Each line in the comb spectrum represents a coherent laser-type emission. 
Filtering out of the high-frequency component through optical demodulation
yields the low-frequency coherent signal as a new promising
type of coherent
terahertz emitters. The EP self-induced oscillation is also a novel
mechanism of optical frequency comb generation as compared to mode-locked lasers
\cite{Udem2002,Cundiff2003} and optical microresonators
\cite{DelHaye2007,Kippenberg2007}.

Consider two coupled EP condensates with order
parameters
\begin{equation}
 \psi_{1,2} = \sqrt{n_{1,2}} e^{i(\Phi \mp \phi)} ,
\label{orderparameter}
\end{equation}
where $n_{1,2}$ are the occupations of the two condensates, $\Phi$ is the
total phase and $2\phi$ is the phase difference. The time evolution of $\psi_{1,2}$ is
governed by the Langevin equations ($\hbar=1$) \cite{Aleiner2012}
\begin{multline}
  \frac{d\psi_\mu}{dt} = -\frac{1}{2}(g\psi_\mu + \gamma\psi_\nu) \\
  -\frac{\mathrm{i}}{2} (2\omega_\mu\psi_\mu - J\psi_\nu + \alpha|\psi_\mu|^2\psi_\mu) + f_\mu(t),
 \label{eq:motion}
\end{multline}
where $\mu \neq \nu =1,2$ label the condensates . The parameter
$g=\Gamma-W$ describes the difference between the rates of losses $\Gamma$ and pumping
$W$, $\omega_\mu$ denote the
singe-particle energies of the condensates, the parameters $\gamma$ and $J$ define dissipative and coherent
coupling between the condensates, respectively, and $\alpha$ is the polariton-polariton
interaction constant. The last term in Eq.\ \eqref{eq:motion} is the Gaussian white noise
satisfying $\langle f_\mu(t)f_{\mu'} \rangle = 0$ and $\langle f_\mu(t) f_{\mu'}^*(t')
\rangle = W_\mu \delta_{\mu\mu'} \delta(t - t')$.
Due to gauge invariance, only the frequency detuning $\omega$ is relevant and in what
follows we will count the frequency from $\omega_0=(\omega_1+\omega_2)/2$. Rescaling time
we can fix $\gamma=1$ and, since rescaling the condensate amplitudes is equivalent to a
change of $\alpha$, we can set $\alpha=2$ without loss of generality.

The dissipative coupling induces a relative phase $\varphi$ dependent  dissipation in the system.
This can be observed
from the eigenvalues $\lambda$ which control the condensate evolution $\psi_{1,2}\sim
{\rm e}^{ \lambda t}$ in the absence of interaction \cite{Aleiner2012}. 
%It follows that $2\lambda_{\pm} = -g
%+\mathrm{i}(\omega_1+\omega_2) \pm \sqrt{(\gamma-\mathrm{i}J)^2 - \omega^2}$, where
%$\omega=\omega_1-\omega_2$. 
%While $\lambda_-$ keeps a negative real part, $\lambda_+$ 
%changes the sign of its real part for sufficiently large $\gamma/g$.
%With increasing pumping, when $g$ becomes small enough, one
%of the two eigenvalues keeps a negative real part and the corresponding eigenmode is
%damped out. The second eigenvalue, however, changes the sign of its real part from
%negative to positive and corresponding eigenmode is pumped in. 
For sufficiently large $\gamma/g$ one of the eigenmodes turns unstable.  Therefore, the dissipative
coupling acts as a phase-selective pump which depends on the relative phase $\varphi$: it
pumps one eigenmode while keeping the other
one lossy. In this regime 
%both the trivial solution $\psi_{1,2}=0$ and the pump-in
%eigenmode are unstable and 
nontrivial weak lasing states are formed (see Ref.\
\cite{Aleiner2012} for a complete account). 
%We note that nontrivial solutions with finite
%occupation numbers exist in the weak lasing domain even without any dissipative
%nonlinearities in Eqs.\ \eqref{eq:motion}. Dissipative nonlinear terms, while
%indispensable in the theory of ordinary lasers, are not necessary in the region
%$g=\Gamma-W>0$ and they can be neglected as compared to the coherent nonlinearities that
%appear due to strong polariton-polariton repulsion.

\begin{figure}
 \centering
 \includegraphics[width=.99\columnwidth]{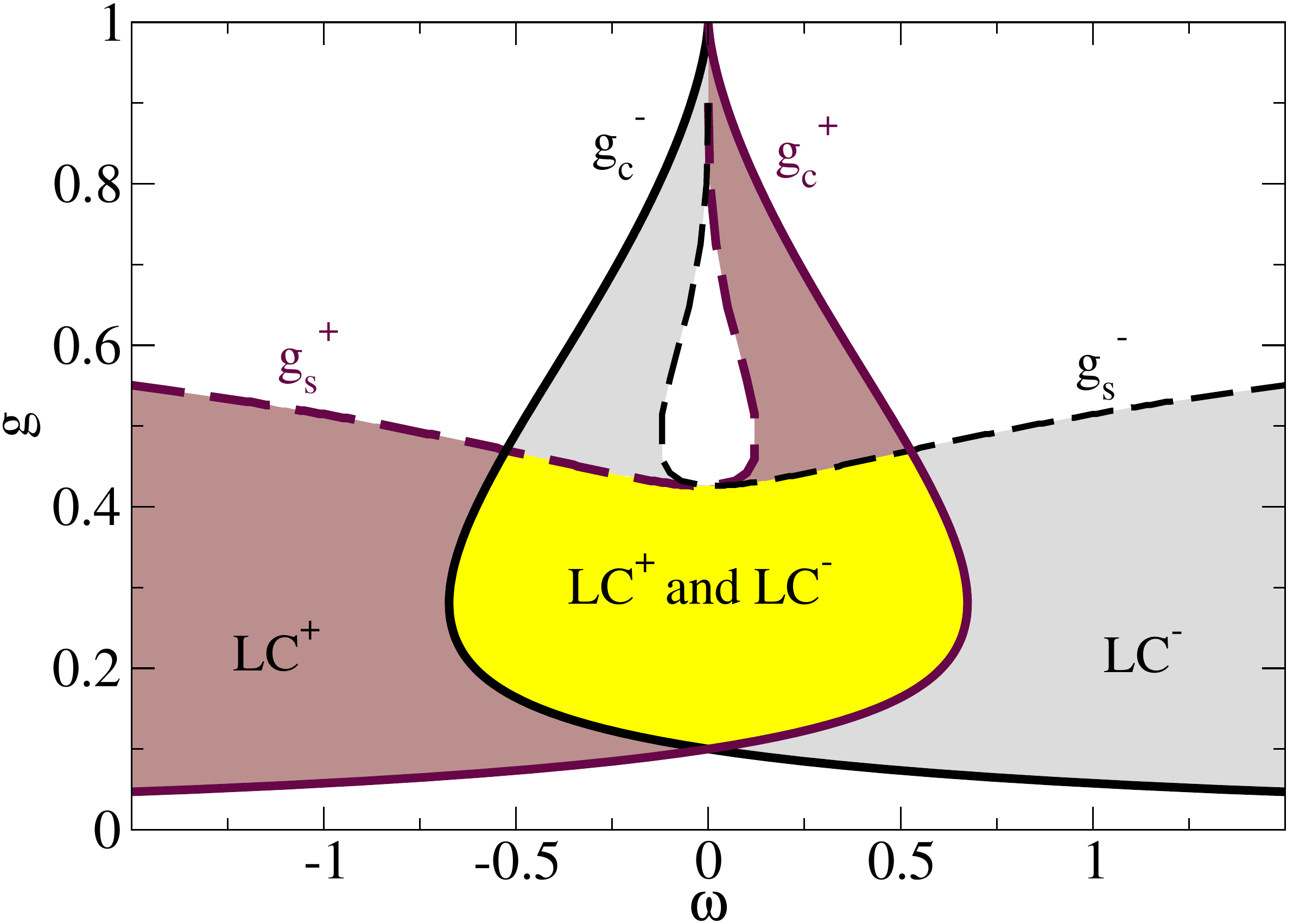}
 \caption{ (Color online)
 Limit cycles LC$^\pm$ appear in the shaded areas in the $\omega, g$-parameter space.
 Stable LC$^\pm$ are born through Hopf bifurcations at the solid $g_c^\pm$ line and turn
 unstable at the dashed $g_s^\pm$ lines where they undergo period doubling bifurcations.
 They are the only stable attractors to coexist in the central (yellow) region.
 Here $J=0.1\gamma$. }
 \label{Fig1}
\end{figure}

First we consider
the noise-free case ($f_{1,2}=0$) in order to determine the attractors of the system
\eqref{eq:motion}.
Two nontrivial solutions F$^{\pm}$ to Eqs.\ \eqref{eq:motion} were identified in Ref.\
\cite{Aleiner2012}. They are characterized by nonzero time-independent triplets
$\{n_1,n_2,\varphi\}_{\pm}$ with the total occupation
$n=n_1+n_2=[(g^2+J^2)R^{\pm}+g\omega]/gJR^{\pm}$, $n_1=n(1-JR^{\pm})/2$,
$n_2=n(1+JR^{\pm})/2$, $\phi=(\pi-\arctan R^{\pm})/2$, where $R^{\pm} = \pm
\sqrt{(1-g^2)/(g^2+J^2)}$. The total phase $\Phi$ of the condensates satisfies
\begin{equation}
 \dot{\Phi} = \left( - \frac{1}{2} + \frac{J\cos(2\phi)}{4\sqrt{n_1 n_2}} \right) (n_1 + n_2)
 + \frac{\sin(2\phi)}{4\sqrt{n_1 n_2}} (n_1 - n_2),
 \label{eq:totalphase}
\end{equation}
and the rhs in (\ref{eq:totalphase}) gives a time-independent frequency
$\dot{\Phi}=-\Omega_0$ for these solutions. The two centers evolve in a coherent fashion
$\psi_{1,2} \sim {\rm e}^{-i\Omega_0 t}$  and $\Omega_0$ defines the blue-shift of the
emission line with respect to the average single-particle frequency. 
In the subspace $\{
n_1,n_2,\phi \}_{\pm}$ these states correspond to fixed points \cite{NoteCycle4D}. 
In parts of the
control parameter space these states are stable, and 
should manifest themselves as weak lasing states. 
The $F^{\pm}$
states loose stability at $g=g_c^{\pm}$ \cite{Aleiner2012}
\begin{equation}
 2(g_c^4 + J^2) R^\pm = (g_c^2 + 1) \left[\omega g_c + (g_c^2 + J^2) R^\pm \right]
\end{equation}
We plot the two instability curves in the $\{g,\omega\}$ space at fixed $J=0.1$ in
Fig.~\ref{Fig1} (solid lines). The $F^{\pm}$ states are unstable in the shaded areas
LC$^{\pm}$. In particular they are both unstable in the joint area LC$^+$ and LC$^-$,
where the trivial solution $n_1=n_2=0$ is unstable as well. What are then the stable stationary states
of the system, if any?

The answer is obtained by linearizing the phase
space flow around $F^{\pm}$ in the subspace $\{ n_1,n_2,\phi \}_{\pm}$. At $g = g_c^\pm$
two corresponding eigenvalues are purely imaginary $\pm i \Delta\Omega$, with their
real parts changing sign.
As a result, a supercritical Hopf bifurcation
occurs, where stable limit cycles LC$^\pm$ with frequency
\begin{equation}
 \Delta\Omega = \sqrt{2g^2 + J^2/g^2 + J^2 + \omega g/R^\pm }
\end{equation}
are born around the respective unstable fixed points F$^\pm$ \cite{NoteTorus,Alexeeva2014}. 

Away from the bifurcation line the LCs increase the oscillation amplitudes, deform, and change their frequency.
The coexistence region of LC$^+$ and LC$^-$ grows in size as the Josephson tunneling is
reduced. At the Hopf bifurcation, where a LC emerges, $n_{1,2}, \varphi$ and also
$\dot{\Phi}$ become periodic functions of time with period $T=2\pi/\Delta\Omega$. Then
they may be expanded in a Fourier series with frequency harmonics $N\Delta\Omega$ and
$N=0,\pm1,\pm2,...$ The integration of the constant term $(N=0)$ in the Fourier series of
$\dot{\Phi}$ results in a linear time dependence, $\Phi_{DC} = - \Omega_0 t$, similar to
$F^{\pm}$. Therefore
\begin{equation}
 \psi_\mu(t) = p_\mu(t) e^{-i \Omega_0 t},
 \label{eq:znotcirclezero}
\end{equation}
where the functions $p_\mu(t) = p_\mu(t + T)$ are periodic in time. The Fourier spectrum
of $\psi_\mu(t)$ is equidistant with frequency harmonics positioned at $\Omega_0 +
N\Delta\Omega$.

%Since the LC evolves in a three-dimensional phase space, its stability is characterized
%by a three-dimensional Floquet matrix with real elements. One eigenvalue is always fixed
%at unity and corresponds to perturbations along the limit cycle. The two remaining
%eigenvalues are either both real, or complex conjugated. An instability occurs when at
%east on of these eigenvalues acquires an absolute value larger than unity. We trace the
%limit cycles numerically with Newton routines, and compute the Floquet eigenvalues.
Approaching the dashed lines $g_s^{\pm}$ in Fig.\ref{Fig1}, 
the corresponding LC turns unstable and undergoes a period doubling bifurcation.
%
%the two eigenvalues first
%collide on the negative real axis inside the unit circle. One of them subsequently
%approaches the unit circle at $-1$ and exits it, leading to a period doubling bifurcation
%precisely at $g_s^{\pm}$ in Fig.\ref{Fig1}. 
This gives rise to a new stable
period-doubled LC, which however again quickly undergoes a period doubling bifurcation. A
period doubling route to chaos along a Feigenbaum scenario leads to chaotic attractors
\cite{Alexeeva2014}. Therefore just two coupled exciton-polariton condensates suffice to
produce an extremely rich and complex synchronized dynamics. 
%This happens because the
%cavity light field participates in the formation of the condensate, and also mediates a
%phase-sensitive interaction between several condensate centers, in addition to the
%expected Josephson coupling which is a straightforward path of exchanging condensate
%particles between the centers.

Experimentally the polariton order parameter is detected by
analyzing the emitted light from the microcavity. In near-field measurements, only small
parts of the sample, like one condensation center, can be probed. Our aim is to calculate
the spectral density $I_{1,2}(\Omega)$ of the radiation corresponding to different
nontrivial attractors. Applying a Fourier transformation (FT) we have
\begin{equation}
I_\mu(\Omega) = |\mathrm{FT}(z_\mu(t))|^2, \qquad \mu = 1,2.
\label{eq:generalintensitynearfield}
\end{equation}

In the fixed points $F^{\pm}$ $\dot{n}_{1,2} =  \dot\phi = 0$, and the time
dependence comes from the evolution of the total phase $\Phi =  - \Omega_0 t$,
\begin{equation}
 \psi_\mu(t) = C_\mu e^{-i \Omega_0 t},
 \label{eq:zforFP}
\end{equation}
with constants $C_\mu$. Thus the condensates emit light at frequency $\Omega_0$ fully synchronized. 
The emission spectrum consists of only one peak, in contrast to the
case of noninteracting polaritons, where two separated peaks are expected.

\begin{figure}
 \centering
 \includegraphics[width=.99\columnwidth]{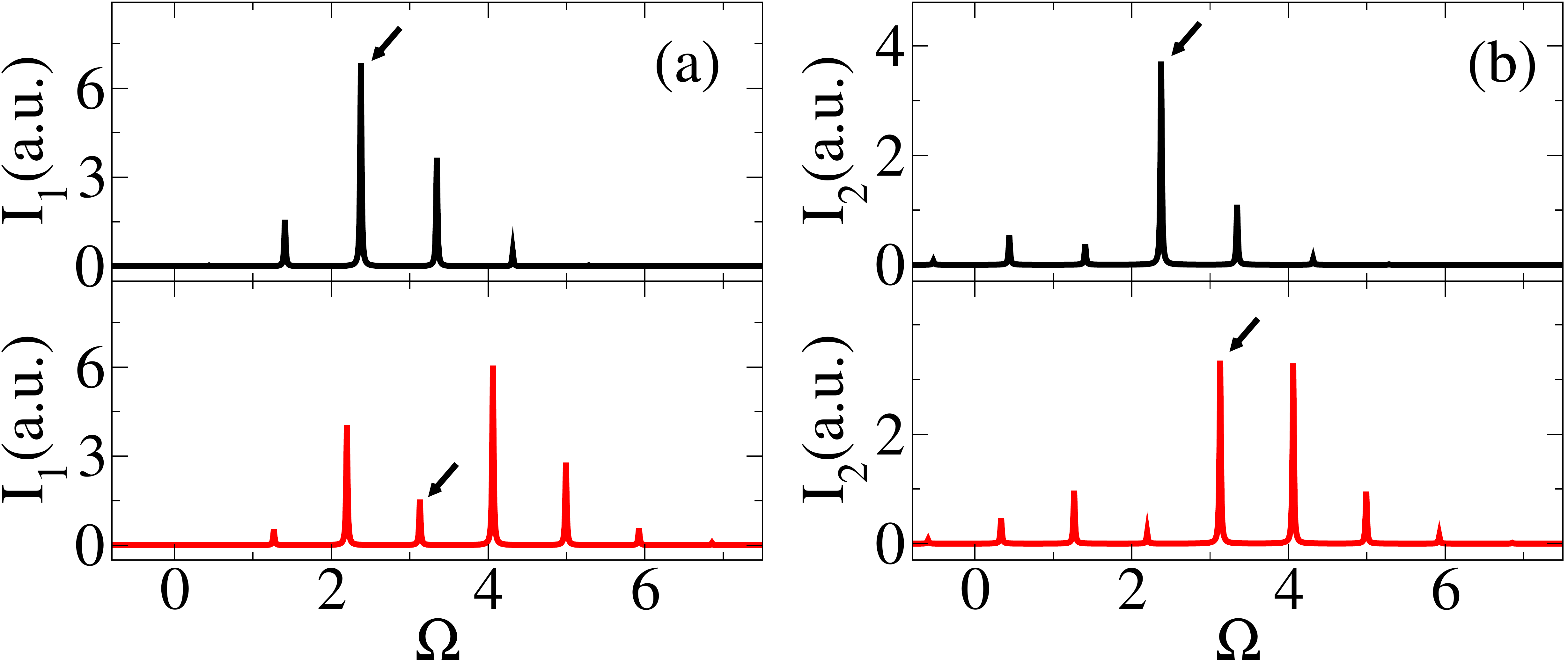}
 \includegraphics[width=.9\columnwidth]{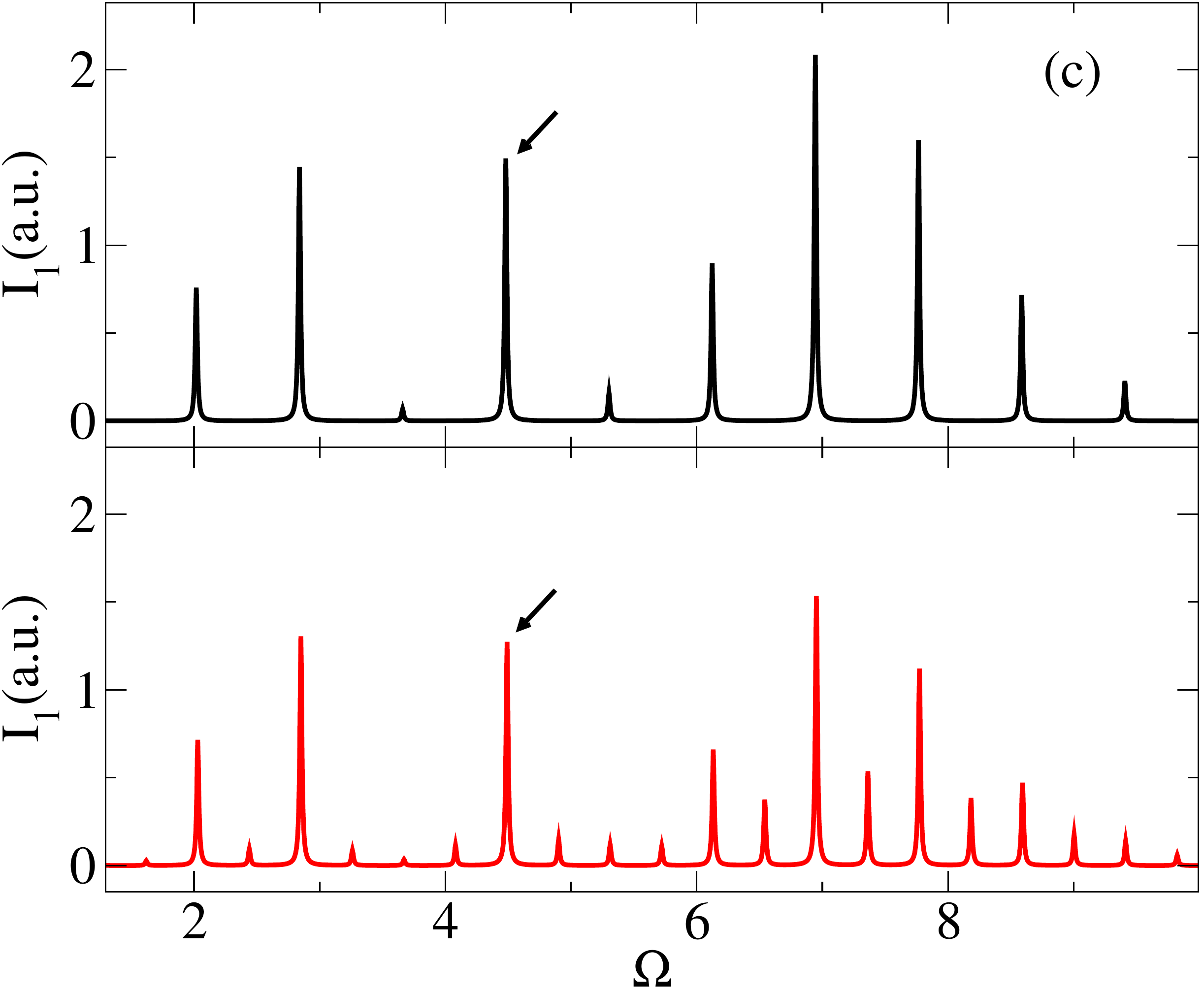}
 \caption{ (Color online)
 Asymmetric frequency combs in the near-field spectrum from a LC$^+$ (a),(b) and
 a period doubled LC$^+$ (c). Here $\omega=0$, $J=0.1$.
 The small arrows indicate the position of the respective N=0 peak.
 (a) $I_1$ with $g=0.19$ (upper panel), $g=0.27$ (lower panel).
 (b) The same as (a) but for $I_2$.
 (c) $I_1$ for $g=0.425$ (upper panel) and $g=0.426$ (lower panel, period doubled LC spectrum).
 For visualization, a small artificial Lorentzian line width was added to all emission lines.  }
 \label{Fig2}
\end{figure}

Since the limit cycles LC$^{\pm}$ are characterized by an equidistant spectrum, we first
numerically compute the corresponding frequency positions, and then calculate the intensity of each
frequency harmonics using a Fourier series expansion. The resulting spectra are shown in Fig.\ref{Fig2}(a,b). Close to the
Hopf bifurcation, there is only one considerable emission peak originating from the $F^+$
spectral line (Fig.\ref{Fig2}(a,b) upper panels). Further away from the Hopf bifurcation,
the satellite peaks grow to form a frequency comb with asymmetric
tails (Fig.\ref{Fig2}(a,b) lower panels). The comb also acquires several
peak maxima, with the highest peak originating from a satellite with nonzero $N=2$
(Fig.\ref{Fig2}(a,b) lower panels). When the LC undergoes a period
doubling bifurcation, the comb becomes twice as dense (Fig.\ref{Fig2}(c)).

The typical modulation frequency is independent of the polariton-polariton interaction
constant $\alpha$ and is of the order of the coupling constant
$\Delta\Omega\sim\gamma$. For typical dissipative rates in semiconductor microcavities
$\Gamma\sim 10^{12}\;\mathrm{ps}^{-1}$ one expects condensate
pairs with dissipative coupling $\gamma\sim 5\times 10^{11}\;\mathrm{ps}^{-1}$ and
below. This would permit to generate frequency combs with terahertz separation
between the individual peaks. Additional reduction of this separation by period doubling
can shift the modulation frequency into the millimeter range.

Finally, we consider the influence of noise in Eq.\
\eqref{eq:motion}. In general, it will broaden the peaks discussed so far, and can lead
to a merging of peaks with small enough spacing.
The emission spectrum can be obtained using the Wiener-Khinchin theorem,
\begin{equation}
 I_\mu(\Omega) = \frac{1}{\pi} \Re \int \limits_0^\infty
 \langle \psi_\mu(t) \psi_\mu^*(0) \rangle e^{i \Omega t} dt,
 \label{eq:lineshape}
\end{equation}
where $\langle \psi_\mu(t) \psi_\mu^*(0) \rangle$ is the auto-correlation of the, now,
random process $\psi_\mu(t)$.

The $F^{\pm}$ states are periodic orbits in the full four-dimensional phase space, and
the dynamics along these periodic orbits is
parameterized by the total phase $\Phi$. While fluctuations off the periodic
orbit will relax back, fluctuations along the orbit do not, and will enforce diffusion of
$\Phi$ on the orbit. The latter fluctuations can be shown to form a Lorentzian line
with the full width at half maximum (FWHM) given by $W(n_1+n_2)/8n_1n_2$ \cite{supplement}. Note that the
FWHM is inversely proportional to the number of particles in the condensate, as it should
be for a laser.

In contrast to the $F^{\pm}$ states, the LC$^\pm$ states are formed by the motion on a torus in the full phase space.
The stability of the attractor demands that fluctuations off the torus relax back.
Fluctuations along the torus surface enforce a diffusion on it. The two nontrivial phases which diffuse, are the total phase $\Phi$ and the second phase
angle, which characterizes the position on the limit cycle.

Close to the Hopf bifurcation, and
in the presence of only a few satellite peaks, we can obtain a closed formula for the
line width. To parameterize the LC we introduce two time arguments: one originating from
the total phase and the other from the LC phase.  Noise in these time
arguments, according to Eq.\ \eqref{eq:znotcirclezero}, leads to 
\begin{equation}
 \begin{split}
  \psi_\mu (t) &= g_\mu \left( t + \frac{1}{-v(t)}
        \int P(\tau) d\tau \right) e^{- i \Omega_0 t +i \int F(\tau) d\tau } \\
        &= \sum \limits_N C_\mu^N e^{-i N\Delta \left( t + \frac{1}{-v(t)} \int P(\tau) d\tau \right) }
        e^{-i \Omega_0 t} e^{i \int F(\tau) d\tau},
 \end{split}
 \label{eq:LCnoise}
\end{equation}
where the periodic function $p(t)$ has been expanded in a Fourier series with
coefficients $C^N$ and $v(t)$ is the velocity of the noise-free trajectory along the LC
in the three-dimensional space $\{n_1,n_2,\varphi\}$. The noise term $P(t)$ is the
projected noise along the LC, while $F(t) = (1/4\mathrm{i})\left[\sum_\mu
(f_\mu(t)/\psi_\mu) - \mathrm{c.c.}\right]$ is the noise added to the rhs of Eq.\
\eqref{eq:totalphase} for $\dot\Phi$. Note that perturbing the time argument of $p(t)$
already accounts for a part of the noise $F(t)$, because $\Phi(t)$  is time periodic.
However close to the Hopf bifurcation this periodic part is negligible compared to the DC
part of $\Phi(t)$, so that the above separation of noise in the time arguments is
justified.

The choice of the two time arguments is convenient when characterizing the dynamics of
the noise-free LC. In the presence of noise however it leads to nonzero correlations
\begin{equation}
 \langle (N P(t) + F(t)) (N P(t') + F(t')) \rangle = 2 \kappa_N(t) \delta(t - t').
 \label{eq:kappaPF}
\end{equation}
Denoting $\langle P(t) P(t')\rangle=\kappa_{PP}\delta(t-t')$,
         $\langle F(t) F(t')\rangle=\kappa_{FF}\delta(t-t')$,
         $\langle P(t) F(t')\rangle=\kappa_{PF}\delta(t-t')$,
we have
\begin{equation}
\kappa_N = \kappa_{FF} +2 N \kappa_{PF} + N^2 \kappa_{PP}
\label{asymmetry}
\end{equation}
with an asymmetric dependence of $\kappa_N$ on $N$.

The intensity of the noise $\kappa_N(t)$ is periodic in time, which
originates from the oscillation of occupation numbers and relative phase for evolution
along the LC. Experimentally, the measurement time spans many LC periods and one can use
the average value $\bar{\kappa}_N = \frac{1}{T} \int \limits_0^T \kappa_N(\tau) d\tau$.
Then
\begin{equation}
 I_\mu(\Omega) = \frac{1}{\pi T} \sum \limits_N \left| C_\mu^N \right| ^2
 \frac{\bar{\kappa}_N}{\bar{\kappa}_N^2 + \left(\Omega-\Omega_0 - N\Delta\Omega\right)^2} .
 \label{eq:finallineshapenoise2}
\end{equation}
We obtain a Lorentzian for every emission line, with an $N$-dependent width
$\bar{\kappa}_N$ according to Eq.\ \eqref{eq:kappaPF}. The $N$-dependence of the line
broadening $\bar{\kappa}_N$ follows from Eq.(\ref{asymmetry}) and shows two remarkable features. First, there is an
$N$-symmetric line broadening with increasing $N$ which is very strong and proportional
to $N^2$. Second, there is an asymmetric contribution $\sim N$ which originates form
nonzero correlations $\kappa_{PF}$. It may lead to a satellite peak becoming more narrow
than the main peak $N=0$, and can further enhance the asymmetry of the spectrum, as
compared to the noise-free case.

\begin{figure}
 \centering
 \includegraphics[width=.99\columnwidth]{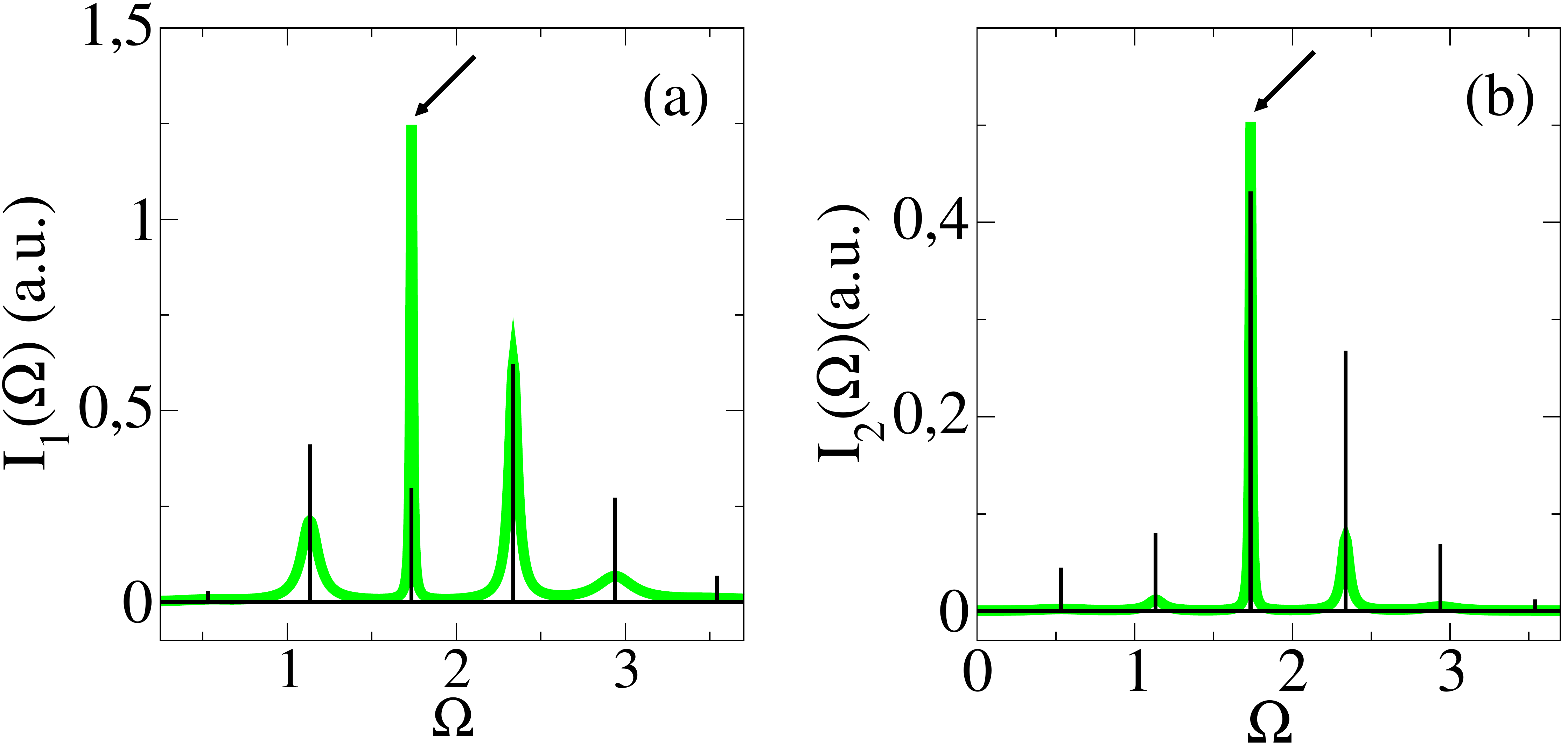}
 \caption{ (Color online)
 Noisy (green lines) against noise-free (black frequency combs) spectrum of a LC$^+$ at
 $g=0.7, \omega=0, J=0.5$, for (a) condensation center 1 and (b) condensation center 2.
 The arrows indicate the central peak. Here $W = 0.02$.}
 \label{Fig3}
\end{figure}

To calculate the line width $\bar\kappa$ numerically, we denote by $(A(t), B(t),
C(t))^\mathrm{T}$ the normalized tangent vector along the LC in the coordinates
$(\phi,n_1,n_2)^\mathrm{T}$. Then $P$ is the noise in $\dot\phi,\dot{n}_1,\dot{n}_2$
projected onto this tangent and we can evaluate from Eq.\ \eqref{eq:kappaPF}
\begin{multline}
  \kappa_N = \frac{W_1}{n_1} \left( N^2 \left(
                \frac{A^2}{4} + B^2 n_1^2 \right) - \frac{NA}{4} + \frac{1}{16}
                                               \right) \\
                   + \frac{W_2}{n_2} \left( N^2 \left(
                   \frac{A^2}{4} + C^2 n_2^2 \right) + \frac{NA}{4} + \frac{1}{16}
                                        \right).
 \label{eq:kappaPFexplicitly}
\end{multline}
We show the spectrum with inhomogeneous line broadening compared to the noise-free case
in Fig.\ref{Fig3}. Due to noise, the asymmetry of the spectrum is enhanced and the
strict equidistance of emission lines is relaxed for strong enough line broadening. We
note that this shape together with the equidistance of emission peaks is very reminiscent
of experimentally obtained spectra in \cite{Krizhanovskii2009}.

Dissipative coupling between coexisting exciton-polariton
condensates in semiconductor microcavities together with strong polariton-polariton
repulsion leads to a rich dissipative nonlinear dynamics already for two coupled
condensates. We showed that, in addition to full synchronization
\cite{Wouters2008,Aleiner2012}, formation of limit cycles gives rise to
frequency combs of equidistant asymmetric spectral lines. The frequency offset and line spacing of the
combs are tunable through the control parameters. Through period doubling, the line
spacing can be additionally reduced by an order of magnitude. This modulated emission can
be useful for terahertz and sub-terahertz applications. 
Shot noise from the pump results in a complex diffusion in phase space and has
strong impact on higher order satellite peaks.

\emph{Acknowledgments.}---We thank I.\ Aleiner, and A.\ Kavokin for
valuable discussions. YGR acknowledges support from the EU FP7 IRSES Project POLAPHEN.

%%%%%%%%%%%%%%%%%%%%%%%%%%%%%%%%%%%%%%%%%%%%%%%%%%%%%%%%%%%%%%%%%%%%%%%%%%%%%%%%%%%%%%

%%%%%%%%%%%%%%%%%%%%%%%%%%%%%%%%%%%%%%%%%%%%%%%%%%%%%%%%%%%%%%%%%%%%%%%%%%%%%%%%%%%%%%
%\bibliographystyle{apsrev.bst}
%\bibliography{DRAFT}

\begin{thebibliography}{99}
\bibitem{NoteCycle4D} Note that F$^\pm$ solutions are limit cycles in the full four-dimensional space
    $\{\psi_1,\psi_2\}$, which becomes essential in the presence of noise.
\bibitem{NoteTorus}
These new solutions are limiting two-dimensional tori in the full four-dimensional space $\{\psi_1,\psi_2\}$, which will become particularly important in the presence of noise.
\end{thebibliography}

\newpage
~
\newpage
\setcounter{equation}{0}
\renewcommand{\theequation}{S\arabic{equation}}
%%%%%%%%%%%%%%%%%%%%%%%%%%%%%%%%%%%%%%%%%%%%%%%%%%%%%%%%%%%%%%%%%%%%%%%%%%%%%
\section{Supplemental Material}

This Supplemental Material presents the derivation of the noise-induced line properties of the $F^{\pm}$ limit cycles.

%\subsection{Noise induced line broadening of the $F^{\pm}$ limit cycles}
%\label{sec:radspecificnoise1}

Let us rewrite the order parameters from  (\ref{orderparameter}) as
\begin{equation}
\psi_{1,2} = \sqrt{n_{1,2}} e^{i\varphi_{1,2}} \;.
\end{equation}
From the equations of motion (\ref{eq:motion}) it is straightforward to find the functions $A,B$ of the differential equations (we omit these details for brevity)
\begin{equation}
\dot{n}_{1,2}= A_{1,2}(n_{1,2},\varphi_{1,2})\;,\; \dot{\varphi}_{1,2}= B_{1,2}(n_{1,2},\varphi_{1,2})\;.
\end{equation}
Adding noise back to the equations of motion (\ref{eq:motion}) yields
\begin{equation}
\dot{\varphi}_{\mu}= B_{1,2}(t) +F_{\mu}(t)\;,\;F_{\mu}=\frac{1}{2i}\left( \frac{f_{\mu}}{\psi_{\mu}} - \frac{f^*_{\mu}}{\psi^*_{\mu}} \right)\;.
\end{equation}
The random process $F_{\mu}(t)$ is characterized by
\begin{equation}
\langle F_{\mu}(t) F_{\mu}(t')\rangle = 2\kappa_{\mu}(t) \delta(t-t')\;,\;\kappa_{\mu}= \frac{W}{4|\psi_{\mu}|^2}.
\end{equation}
In a $F^{\pm}$ limit cycle the noise-free evolution of $\varphi_{1,2}(t)$ is determined by the evolution of the total phase given by $\dot{\Phi} = - \Omega_0$.
The presence of noise gives
\begin{equation}
 \dot{\Phi} = - \Omega_0(t) + F(t)\;,\;F(t)= \frac{1}{2}(F_1(t)+F_2(t))\;.
 \label{eq:motionPhinoise}
\end{equation}
Since $\dot{n}_{1,2} = \dot{\varphi} = 0$ we obtain
\begin{equation}
\langle F(t) F(t') \rangle = 2 \kappa \delta(t - t'), \kappa = \frac{W}{16} \left[ \frac{1}{n_1} + \frac{1}{n_2} \right] .
\label{eq:commonFcorrelation}
\end{equation}
The emission line is finally given by
\begin{equation}
 \begin{split}
  I_\mu(\Omega) &= \frac{n_{\mu}}{\pi} \Re \int \limits_0^\infty dt \langle e^{i \int \limits_0^t F(\tau) d\tau} e^{i \left( \Omega - \Omega_0 \right) t} \rangle \\
                 &= \frac{n_{\mu}}{\pi} \Re \int \limits_0^\infty dt \langle e^{- \kappa t} e^{i \left( \Omega - \Omega_0 \right) t} \rangle \\
                 &= \frac{n_{\mu}}{\pi} \frac{\kappa}{\kappa^2 + (\Omega - \Omega_0)^2} ,
 \end{split}
 \label{eq:lineshapefixedpoint}
\end{equation}
which is a Lorentzian centred at $\Omega_S$ and of width $\kappa$.

\end{document}